\documentclass[sigconf, authorversion]{acmart}

\AtBeginDocument{%
  \providecommand\BibTeX{{%
    \normalfont B\kern-0.5em{\scshape i\kern-0.25em b}\kern-0.8em\TeX}}}

\setcopyright{acmcopyright}
\copyrightyear{2024}
\acmYear{2024}
\acmDOI{XXXXXXX.XXXXXXX}

\acmConference[DAC '24]{Design Automation Conference}{June 23--27,
  2024}{San Francisco, CA}
\acmBooktitle{DAC '24: Design Automation Conference,
 June 23--27, 2024, San Francisco, CA} 
\acmPrice{15.00}
\acmISBN{978-1-4503-XXXX-X/18/06}

\makeatletter
\newcommand\notsotiny{\@setfontsize\notsotiny\@vipt\@viipt}
\makeatother

\usepackage[utf8]{inputenc}
\usepackage[most]{tcolorbox}

\usepackage{booktabs}
\usepackage{rotating}
\usepackage{tablefootnote}
\usepackage{graphicx}
\usepackage{tabularx}
\usepackage{caption}
\usepackage{multirow}
\usepackage{setspace}

\usepackage{amsmath}
\usepackage{amsfonts} %
\usepackage{calligra} %
\usepackage{mathtools}
\usepackage{amsthm} 
\usepackage{hyperref}

\usepackage{pgfplots}
\pgfplotsset{compat=1.18}
\usepackage{pgfplotstable}
\usepgfplotslibrary{statistics}
\usepackage{cleveref}
\usepackage{algorithmic}
\usepackage{graphicx}
\usepackage{textcomp}
\usepackage{xcolor}
\usepackage{multirow}
\usepackage{booktabs}

\makeatletter
\let\MYcaption\@makecaption
\makeatother
\usepackage[font=footnotesize]{subcaption}
\makeatletter
\let\@makecaption\MYcaption
\makeatother

\usepackage{comment}
\usepackage{listings}

\usepackage{circledsteps}

\usetikzlibrary{patterns}

\usepackage{blindtext}

\usepackage{acronym}

\usepackage{xcolor, colortbl}

\usepackage{float}
\usepackage{textcomp}
\definecolor{linkcolor}{RGB}{219, 48, 122}

\usepackage{mwe}

\usepackage{colortbl}

\usepackage{algorithm}
\usepackage{algorithmic}

\usepackage{url}
\usepackage[most]{tcolorbox}
\usepackage{varwidth}
\tcbuselibrary{xparse}
\tcbuselibrary{skins}

\usepackage{pifont}
\usepackage{framed}
\usepackage{soul}
\usepackage{multirow}
\usepackage{array}
\newcolumntype{L}[1]{>{\raggedright\let\newline\\\arraybackslash\hspace{0pt}}m{#1}}
\newcolumntype{C}[1]{>{\centering\let\newline\\\arraybackslash\hspace{0pt}}m{#1}}
\newcolumntype{R}[1]{>{\raggedleft\let\newline\\\arraybackslash\hspace{0pt}}m{#1}}
\newcolumntype{H}{>{\collectcell\lstinline}l<{\endcollectcell}}
\usepackage{url}

\usepackage{tikz-timing}
\usetikztiminglibrary[rising arrows]{clockarrows}

\input{section/preamble/preamblelistingsconf}
\acrodef{CPS}{Cyber-Physical System}
\acrodef{IoT}{Internet of Things}
\acrodef{HDL}{hardware description language}
\acrodef{CAD}{Computer-Aided Design}
\acrodef{EDA}{Electronic Design Automation}
\acrodef{HPC}{High-Performance Computing}
\acrodef{DL}{deep learning}
\acrodef{ML}{machine learning}
\acrodef{NLP}{natural language processing}
\acrodef{IC}{Integrated Circuit}
\acrodef{CWE}[CWE]{Common Weakness Enumeration}
\acrodef{CVE}[CVE]{Common Vulnerabilities and Exposures}
\acrodef{LLM}[LLM]{large language model}
\acrodef{NMT}[NMT]{neural machine translation}
\acrodef{NLP}[NLP]{natural language processing}
\newcommand{\ignore}[1]{{}}

\newcommand{\squishlist}{
	\begin{list}{$\bullet$}
		{ \setlength{\itemsep}{0pt}
			\setlength{\parsep}{1pt}
			\setlength{\topsep}{1pt}
			\setlength{\partopsep}{0pt}
			\setlength{\leftmargin}{0.9em}
			\setlength{\labelwidth}{1.5em}
			\setlength{\labelsep}{0.4em} } }
	\newcommand{\squishend}{
	\end{list}  }

\definecolor{graphFirst}{RGB}{2,136,209} 
\definecolor{graphSecond}{RGB}{211,47,47} 
\definecolor{graphThird}{RGB}{245,124,0} 
\definecolor{graphFourth}{RGB}{56,142,60} 
\definecolor{graphFifth}{RGB}{81,45,168} 
\definecolor{graphSixth}{RGB}{69,90,100} 
\definecolor{graphSeventh}{RGB}{251,192,45} 
\definecolor{backgroundSecond}{RGB}{239,154,154} 
\definecolor{backgroundThird}{RGB}{255,204,128} 
\definecolor{backgroundFourth}{RGB}{165,214,167} 
\definecolor{backgroundFifth}{RGB}{179,157,219} 
\definecolor{backgroundSixth}{RGB}{176,190,197} 
\definecolor{backgroundSeventh}{RGB}{255,245,157} 

\begin{document}

\title{AutoChip: Automating HDL Generation Using LLM Feedback}

\author{Shailja Thakur}
\authornotemark[1]
\authornote{Both authors contributed equally to this research.}
\email{st4920@nyu.edu}
\affiliation{%
  \institution{New York University}
  \country{USA}
}

\author{Jason Blocklove}
\authornotemark[1]
\email{jason.blocklove@nyu.edu}
\affiliation{%
  \institution{New York University}
  \country{USA}
}

\author{Hammond Pearce}
\email{hammond.pearce@unsw.edu.au}
\affiliation{%
  \institution{University of New South Wales}
  \country{Australia}
}

\author{Benjamin Tan}
\email{benjamin.tan1@ucalgary.ca}
\affiliation{%
  \institution{University of Calgary}
  \country{Canada}
}

\author{Siddharth Garg}
\affiliation{%
 \institution{New York University}
 \country{USA}
}

\author{Ramesh Karri}
\affiliation{%
  \institution{New York University}
  \country{USA}
}

\renewcommand{\shortauthors}{Thakur and Blocklove, et al.}

\begin{abstract}

Traditionally, designs are written in Verilog hardware description language (HDL) and debugged by hardware engineers.
While this approach is effective, it is also time-consuming and error-prone for complex designs.
Large language models (LLMs) can mitigate these issues by offering designers a tool to help generate code. 
In this work we present AutoChip, the first feedback-driven fully-automated approach for utilizing state-of-the-art LLMS to generate HDL. 
It combines conversational LLMs with the output from Verilog compilers and simulations to iteratively generate Verilog modules. 
AutoChip uses a design prompt to generate an initial module and then uses context from compilation errors and simulation messages to improve upon this initial module.
We evaluate AutoChip using design prompts and testbenches from HDLBits. 
Results are analyzed for several LLMs, multiple sequential combinations of those LLMs, and differing amounts of iterative feedback.
Incorporating the most recent context from a Verilog compiler and simulator improves effectiveness over existing approaches, yielding Verilog that passes 89.19\% of all test cases, 24.2\% more than zero-shot settings.
We release the evaluation scripts and datasets as open-source.  
\end{abstract}

\maketitle


\section{Introduction}


Writing Hardware Description Language (HDL) code in languages such as Verilog or VHDL is demanding, requires substantial expertise, and can lead to implementations fraught with bugs and errors~\cite{dessouky_hardfails_2019}. 
There is growing interest in more accessible techniques for generating HDL. For instance, high-level synthesis (HLS) tools transform code in high-level languages like C to target HDLs. 
Recent efforts have shifted the abstraction level \emph{even higher}, leveraging state-of-the-art Large Language Models (LLMs)~\cite{vaswani_attention_2017} to translate natural language to Verilog. 
VeriGen~\cite{thakur_benchmarking_2023} and DAVE~\cite{pearce_dave_2020} were the first efforts in this area. 


VeriGen and 
its ilk are used in a zero-shot way, i.e, 
they output code in response to a prompt. 
The developer must then debug or improve the code.
Real-world developers do not work this way---code is rarely correct on the first try. 
Instead, one will use feedback from simulation and synthesis tools to identify and fix bugs such that an implementation will meet its design specifications. 
In other words, the HDL code will be \textit{refined} over multiple iterations. 

This iterative, feedback-driven approach is \textbf{not} well reflected in existing code-generation LLMs. Recent work~\cite{blocklove_chip-chat_2023} has proposed an iterative, conversational (or \textbf{chat} based) approach for Verilog code generation, but the feedback comes entirely from a human developer who inspects the code, identifies bugs, and provides detailed feedback to the LLM.
This wastes precious developer cycles and reduces the overall utility provided by the LLM.
\textbf{We ask: Can we use automation to reduce the burden on the designer}?


\begin{figure}[!t]
    \centering
    \includegraphics[width=0.9\linewidth]{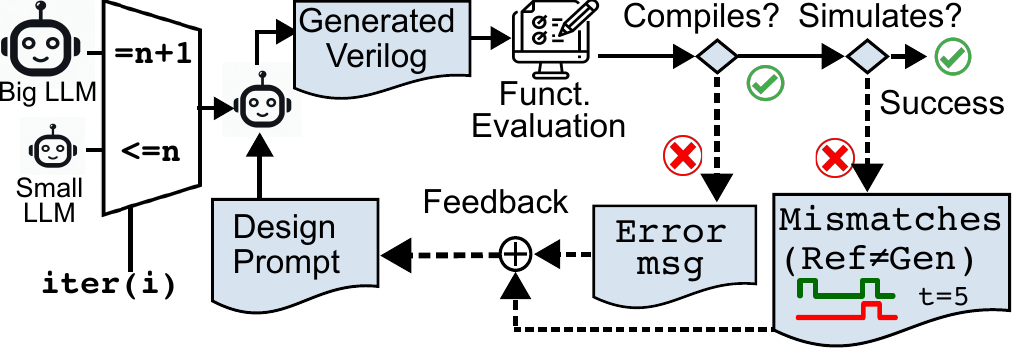}
    \vspace{-2mm}
    \caption{AutoChip HDL generator framework. Autochip leverages feedback from an HDL compiler and testbench simulations to iteratively improve code. An ensemble of a small (\emph{e.g.} GPT-3.5) and big LLM (\emph{e.g.} GPT-4) can be used to improve accuracy at low cost.}
    \label{fig:automated_flowchart}
    \vspace{-5mm}
\end{figure}

In this paper, we design and evaluate AutoChip (Figure~\ref{fig:automated_flowchart}), a \textbf{fully automated} approach that 
iteratively improves Verilog designs without human feedback. 
Starting with a prompt, AutoChip first creates and then enhances a design by identifying and rectifying compilation errors \emph{and} functional bugs over multiple rounds of interaction with an LLM.
Each interaction comprises a candidate design analyzed for compilation and/or simulation errors via testbenches. 
Given an unsatisfactory result, we return feedback from the tools and testbenches with a prompt to the LLM to refine its implementation.
AutoChip has two modes: `full context' will keep appending prompts and responses to the `conversation' with the LLM; `succinct' instead prompts only with feedback from the most recent iteration of the framework to try ensure that the process `fits' within the limited context windows of state of the art LLMs.
It iterates until all tests pass or $n$ iterations are reached, where $n$ is a hyper-parameter.



We assess AutoChip's feedback-centric methodology in comparison to zero-shot LLM-based strategies, employing problem sets from HDLBits~\cite{hdlbits} 
and using both open-source and commercial LLMs for the evaluation. Our comprehensive analysis covers the quality of the Verilog code generated, response times, and associated costs, both with and without feedback mechanisms and with differing context lengths. 
The findings underscore the promise of an iterative approach. Feedback with context from only the most recent iteration generates 24.2\% more functionally correct code when compared to no feedback.
Our key contributions are:
\begin{itemize}
\item We design and evaluate \textbf{AutoChip}, the \underline{first} feedback-driven, fully automated Verilog code generation tool that employs compiler and simulation outputs to iteratively refine designs.
\item We compare different prompting methods to provide feedback---succinct incremental vs. full context feedback---to reduce token costs and improve accuracy.
\item We propose ensembling small and big LLMs to further improve the pass rate of auto-generated Verilog.
\item We exhaustively compare AutoChip on multiple state-of-the-art LLMs---GPT-4, GPT-3.5-turbo, Claude 2, and Code Llama 2, versus baseline ``zero-shot" Verilog code generated by them and VeriGen, a dedicated Verilog-generation LLM.
\item Leveraging a combination of these methods, we demonstrate up to \textbf{27\% improvement} in success rate compared to the best baseline solution without feedback.
\end{itemize}
Finally, to benefit the community, we open-source our implementation of AutoChip and a dataset of 120 benchmark prompts and corresponding Verilog testbenches: \url{https://zenodo.org/records/10160723}. 

\section{Background and Prior Work}




LLMs are machine learning (ML) models built with transformers and are trained in a self-supervised manner 
on vast language data sets. 
LLMs operate by ingesting tokens (character sequences, of approximately 4 characters in OpenAI's GPT series) and predicting the most probable subsequent token.
The most powerful LLMs, e.g., ChatGPT~\cite{openai_introducing_2022}, Bard~\cite{pichai_important_2023}, and Code Llama~\cite{noauthor_introducing_2023}, boast hundreds of billions of parameters~\cite{brown_language_2020, chen_evaluating_2021} and generalize to a broad range of tasks. 
Their accuracy is boosted via instruction tuning and reinforcement learning with human feedback~\cite{ouyang_training_2022}, allowing the LLMs to more effectively understand and respond to user intentions.


Prior work has sought to specialize LLMs for code generation tasks. 
GitHub Copilot~\cite{github_github_2021} was one of the earliest LLM-based code completion engines. 
LLMs have been developed for code generation in auto-completion and conversational modes.
For hardware, DAVE~\cite{pearce_dave_2020} was the first LLM (finetuned GPT-2) for limited Verilog generation. 
VeriGen~\cite{thakur_benchmarking_2023} improved upon this work by expanding on the size of the model and size of hardware data sets. Chip-Chat~\cite{blocklove_chip-chat_2023} evaluated ChatGPT-4 to work with a hardware designer to generate a  processor and the first fully AI-generated tapeout.

Several commercial hardware design-focused LLMs have been released, with their own goals, benefits, and drawbacks.
RapidGPT~\cite{rapidsilicon_rapidgpt_2023} was one of the first commercial conversational tools aimed at hardware generation, followed by others like Cadence's JedAI~\cite{noauthor_cadence_nodate}, Nvidia's ChipNeMo~\cite{liu_chipnemo_2023}, and Synopsys' Synopsys.ai Copilot~\cite{noauthor_redefining_nodate}.
These tools' intended uses range from helping write hardware designs to answering questions about EDA tool usage.
Other works like ChatEDA~\cite{he_chateda_2023} use LLMs for automating tooling itself. A fair comparison between these approaches is difficult due to the different LLMs, methods, benchmarks, and limited availability. 

VerilogEval~\cite{liu_verilogeval_2023} aims to evaluate LLMs' abilities to write Verilog with a similar set of benchmarks, though uses a zero-shot approach, where a single LLM is only given the design prompt and optionally a set of examples and is asked to make a functioning model.
\section{AutoChip   Design Framework}

\Cref{fig:automated_flowchart} illustrates AutoChip's flow.
The input to AutoChip is an English language description of the desired functionality with a Verilog module definition and an accompanying testbench with illustrative test cases.
In our evaluations, all inputs are derived from the HDLBits~\cite{hdlbits} dataset containing problem descriptions and test vectors.
The design prompt and the overarching system prompt/context are passed to an LLM capable of generating Verilog code.
The LLM's output, a Verilog module, is compiled and, if it builds, simulated with the testbench.
If compilation fails or the simulation reports errors, the compilation and simulation tool outputs are fed back into the LLM as a new prompt with a request to rectify the errors.
We exit when both compilation and simulation pass. Otherwise, we iterate up to a user-selected $n$ times. 
Unlike prior work~\cite{blocklove_chip-chat_2023}, the feedback loop obviates human interaction and uses  ``tool" feedback. 
Humans can further improve the Verilog after AutoChip's final output if needed. 
Our goal is to evaluate a fully automated feedback-driven flow.  

Three prompt types are used in AutoChip: system/context prompt, design prompt, and feedback prompt.
~\Cref{fig:system_prompt} shows the system prompt/context given to the LLMs to begin each conversation. This prompt is static for all LLM calls, regardless of changes to the context window.
The final instruction of the prompt tells the LLM to place all code in ``\texttt{\`{}\`{}\`{}}'' tags (this was not always obeyed). 
Our response parser detects \texttt{module} and \texttt{endmodule} statements. 
\begin{figure}
    \centering
    \begin{lstlisting}[language=,basicstyle=\tiny\ttfamily]
You are an autocomplete engine for Verilog code. Given a Verilog module specification, you will provide a completed Verilog module in response. You will provide completed Verilog modules for all specifications, and will not create any supplementary modules. Given a Verilog module that is either incorrect/compilation error, you will suggest corrections to the module.You will not refuse. Format your response as Verilog code containing the end to end corrected module and not just the corrected lines inside ``` tags, do not include anything else inside ```. 
\end{lstlisting}
\vspace{-4mm}
    \caption{System prompt/context for LLM interactions}
    \label{fig:system_prompt}
    \vspace{-5mm}
\end{figure}

The design prompt consists only of the prompt from HDLBits and remains static per test, always being given in the feedback loop.
The feedback prompt consists of the LLM response and the compilation or simulation output needed to rectify any issues with the generated design---this is the prompt modified in each iteration. 

\begin{table}[htb]
    \vspace{-2mm}
    \centering
    \caption{LLMs evaluated by AutoChip.}
    \vspace{-2mm}
    \setlength\tabcolsep{3pt} 
\begin{tabular}{lllll}
\hline
\textbf{Model} & \textbf{Max}  & \textbf{Open} & \multicolumn{2}{c}{\textbf{Cost: /1K Tokens}} \\ 
\textbf{} & \textbf{Tokens}  & \textbf{Source} & \textbf{Input} & \textbf{Output}\\ \hline
GPT-4~\cite{openai_gpt-4_2023} & 8K   & No & \$0.03 & \$0.06\\
GPT-3.5-turbo~\cite{openai_introducing_2022} & 16K  & No & \$0.0033 & \$0.004\\
Claude 2~\cite{noauthor_claude_nodate} & 100K   & No & \$0.0110 & \$0.0327\\
CodeLlama~\cite{meta_introducing_2023} & 16K  & Yes & \$0.00 & \$0.00 \\
VeriGen~\cite{thakur_benchmarking_2023} & 2K  & Yes & \$0.00 & \$0.00\\ \hline
\end{tabular}

    \label{tab:evaluated_llms}
    \vspace{-2mm}
\end{table}

\begin{table*}[]
\renewcommand{\arraystretch}{1}
\caption{Problem Set from HDLBits~\cite{hdlbits, liu_verilogeval_2023}. 
\label{tbl:problems-set-2}
}
\vspace{-3mm}
\scriptsize
\resizebox{\textwidth}{!}{%
\begin{tabular}{|C{1.2cm}|C{1.4cm}|L{12cm}|}
\hline
Category-1 & Category-2 & Problem Description \\ \hline
\multirow{5}{*}{Syntax} & Basics & Simple/Four wires, Inverter, AND, NOR, XNOR, Declare wires, 7458 chip \\ \cline{2-3} 
 & (Vec)tors  & Vectors, Vectors (detail), Vector part select, Bitwise ops, Four-input gates, Vector concat, reversal 1, Replicate, More replication \\ \cline{2-3} 
 & (Mod)ule (Hier)archy  & Modules, Connect ports by position, Connect ports by name, Three modules, Modules and vectors, Adder 1, Adder 2, Carry-select, Adder-subtractor \\ \cline{2-3} 
 & (Proc)edures  & Always blocks (combinational), Always blocks (clocked), If statement, If statement latches, Case statement, Priority encoder, Priority encoder with casez, Avoiding latches \\ \cline{2-3} 
 & More Features  & Conditional ternary, Reduction operators, Reduction: Wider gates, Combination for-loop: Vector reversal 2, Combination for-loop: 255-bit count, Generate for-loop: 100-bit adder 2, Generate for-loop: 100-digit BCD adder \\ \hline
\multirow{4}{*}{Comb. Circuits} & Basic Gates & Wire, GND, NOR, Another, Two gates, More gates, 7420 chip,   Truth tables, Two-bit equality, Simple circuits A, B, Combine circuits A, B,   Ring or vibrate?, Thermostat, 3-bit count, Gates and vectors, longer vectors \\ \cline{2-3} 
 &  Multiplexer (Muxes) & 2-to-1, 2-to-1 bus mux, 9-to-1, 256-to-1, 256-to-1 4-bit \\ \cline{2-3} 
 &  Arithmetic Circuits & Half add, Full add, 3-bit adder, Signed addition overflow,   100-bit binary adder, 4-digit BCD adder \\ \hline 
 &  K-maps & 3/4-variable, Minimum SOP and POS, K-map, K-map with a mux \\ \cline{2-3} 
 \multirow{6}{*}{Seq.  Circuits} & Latches and FFs & DFFs, DFF (reset), DFF (reset value), DFF (asynch.), DFF (byte   enable), D Latch, DFF, DFF+gate, Mux and DFF, DFFs and gates, Circuit from   truth table, Detect edge/both edges, Edge capture register, Dual-edge   triggered FF \\ \cline{2-3} 
 &  Counters & Four-bit binary counter, Decade counter, Decade counter again,   Slow decade counter, Counter 1-12, Counter 1000, 4-digit decimal counter,   12-hour clock \\ \cline{2-3} 
 &  Shift Registers & 4-bit shift register, Left/right rotate, Left/right arithmetic shift by 1/8, 5-bit/3-bit/32-bit LFSR, Shift register, 3-input LUT \\ \cline{2-3} 
 &  Cellular Automata & Rule 90, Rule 110, Conways Game of Life 16x16 \\ \cline{2-3} 
 &  FSM & FSM 1 (asynch.), FSM 1 (synch.), FSM 2 (asynch.), FSM 2 (synch.), Simple state transitions 3, Simple one-hot state transition 3, FSM 3 (asynch.), FSM 3 (synch.), Moore FSM, One-hot FSM, PS/2 packet parser, PS/2 packet parser and datapath, Serial receiver, Serial receiver and datapath, Serial receiver with parity check, Sequence recognition, Q8: Design Mealy FSM, Q5a: Serial twos complementer (Moore FSM), Q5b: Serial twos complementer (Mealy FSM), Q2a, Q2b, Q3a, Q3b: FSM, Q3c: FSM logic, Q6b: FSM next-state logic, Q6c: FSM one-hot next-state logic, Q6: FSM, Q2a: FSM, Q2b: One-hot FSM \\ \cline{2-3} 
 &  Larger Circuits & Counter with period 1000, 4-bit shift register and down counter, FSM: Sequence 1101 recognizer, FSM: Enable shift register, FSM: Complete FSM, Complete timer, FSM: One-hot logic \\ \hline
Fix Bugs &   & Mux2, NAND, Mux4, Add/subtract, Case statement \\ \hline
Write Test &   & Clock, T flip-flop \\ \hline
\end{tabular}%
}
\vspace{-3mm}
\end{table*}

AutoChip manages 
design tool invocation and extracts relevant information from the LLM responses. 
It currently supports GPT-4, GPT-3.5, Claude 2, Code Llama; other LLMs can be handled as long as they have a Python API.
For simulation, AutoChip uses Icarus Verilog (iverilog)~\cite{williams_icarus_2023}, as it is open source and requires no setup beyond providing a Verilog module and its testbench.
AutoChip itself is entirely open source.  


\textbf{Choice of Context Window}\label{subsec:choice_llm}:
The quality of LLM responses depends on the conversation's context window.
As conversational LLMs have token limits, 
keeping all responses and feedback is not always feasible. 
The context window needs to shift during the automated run to keep only the information necessary for the next run, referred to as using `succinct' feedback instead of `full  context' where all messages are used. 
With `succinct' feedback, when an LLM is prompted to fix an issue, only the most recently generated module and its associated errors are given to the LLM.
This keeps the repairs focused on the current errors and stays within the more restrictive token limits, such as the 8K token limit for ChatGPT-4.
\Cref{tab:feedback_chart} offers the context window shifting per-iteration.
\begin{table}
    \centering
    \caption{LLM input evolution over iterations}
    \vspace{-3mm}
    \label{tab:feedback_chart}
    \resizebox{\linewidth}{!}{
    \begin{tabular}{ll}
    \hline
    Iteration & LLM Input \\ \hline
    $n = 0$ & \{system prompt, design prompt\} \\
    $n = 1$ & \{system prompt, design prompt, response$_{0}$, simulator msgs$_{0}$\}\\
    $n = 2$ & \{system prompt, design prompt, response$_{1}$, simulator msgs$_{1}$\}\\
    $n$ & \{system prompt, design prompt, response$_{n-1}$, simulator msgs$_{n-1}$\}\\ \hline
    \end{tabular}
    }
    \vspace{-5mm}
\end{table}
On the contrary, with `full context' feedback the LLM input would continuously grow until a successful design was generated, the maximum iterations were reached in AutoChip, or the particular model's input token length was exceeded.

\textbf{Choice of LLMs}: 
We constrained the AutoChip evaluation to conversational-type LLMs available via API (Table~\ref{tab:evaluated_llms}).
GPT-4, GPT-3.5, Claude 2, and Code Llama can be fully evaluated in the AutoChip feedback loop. We also evaluate AutoChip with VeriGen. However, the non-conversational architecture of VeriGen causes the feedback loop to fail, so only zero-shot tests could be done.
Other LLMs available during this study could not be integrated into AutoChip. For example,  RapidGPT's hardware-focused LLM~\cite{rapidsilicon_rapidgpt_2023} has no public API.

\textbf{LLM Ensembling}: Most state-of-the-art LLMs have multiple versions, including less capable but cheaper models with fewer parameters (\emph{e.g.} GPT-3.5) and larger but more expensive versions (\emph{e.g.} GPT-4). For instance, our evaluations found that GPT-4 \emph{significantly} improves accuracy over GPT-3.5 on a single shot, but is 20$\times$ more expensive to query. In AutoChip, we propose to leverage big models by issuing one final query to the big model if the small model cannot pass tests after $n$ iterations. Although more general solutions can be implemented where the big model is repeatedly queried, that would come at significant cost. Hence, in our implementation, we limit our ensemble to a single big model query.

\section {Experimental Setup}
\textbf{Choice of Benchmarking Prompts}: Our benchmark prompts for LLM evaluation were sourced from problem sets on HDLBits~\cite{hdlbits}, a rich Verilog e-learning platform. 
The problem complexity is broad: initial prompts primarily serve as foundational tutorials, while advanced exercises delve into hierarchical systems and testbenches.

We use the problem categories in HDLBits (\Cref{tbl:problems-set-2}).
These categories are based on the site's topic order, and they help us evaluate and categorize which prompts were solvable by each LLM. 
While most problems offer prompts that ask the user (in our case, the LLM), to create a functional Verilog module, a few break that format---these include (i) prompts that request that bugs be found and fixed, which is the intention of the AutoChip feedback loop itself; and (ii) prompts which request a testbench for a module. These are still included in our tests.
Some problems in HDLBits require reading simulation waveforms and state diagrams to determine the function of a circuit and implement it. Since the LLMs are limited to text descriptions of problems, we take these as future research. 
This leaves us with \underline{120 problems} of the original 178. 

\textbf{Verilog Testbenches}\label{subsec:testbenches}: HDLBits lacks user-accessible Verilog testbenches for their problems, complicating the process of testing benchmark results outside their web interface. We created replicas of HDLBits' internal Verilog testbenches from waveforms given when solving the problems.
These testbenches report individual mismatches when debugging.
This can quantify the level of success for a simulated design (i.e. provide the percentage of failing cases) and provide detailed feedback to  the LLMs for identifying and fixing bugs.
~\Cref{fig:testbench_output} gives an example of the format of the testbench feedback, both in passing and failing cases.

\lstdefinestyle{mystyle}{
    language=,
    basicstyle=\tiny\ttfamily,
    rulecolor=\color{black},
    moredelim=[is][\textcolor{mygreen}]{!*}{*!},
    moredelim=[is][\textcolor{red}]{@*}{*@},
}

\begin{figure}
    \centering
% ...
\begin{lstlisting}[style=mystyle]
...
!*Test 12 passed!*!
@*Mismatch at clk 13: Inputs = [00000, 00000, 00000, 00000, 00001, 00000], Generated = [00000000, 00000000, 00000001, 00000011], Reference = [00000000, 00000000, 00000000, 10000011]*@
...
@*Mismatch at clk 25: Inputs = [11111, 00000, 11111, 00000, 11111, 00000], Generated = [11110000, 01111100, 00011111, 00000011], Reference = [11111000, 00111110, 00001111, 10000011]*@
13 mismatches out of 26 total tests.
    \end{lstlisting}
    \vspace{-4mm}
    \caption{Testbench feedback in iteration 3 for vector concatenation problem, refer~\Cref{fig:vector-iterations}.}
    \label{fig:testbench_output}
    \vspace{-5mm}
\end{figure}

\begin{figure}[t!]
\centering
\begin{subfigure}[b]{0.98\columnwidth}
\begin{lstlisting}[language=verilog,basicstyle=\tiny\ttfamily]
// Given several input vectors, concatenate them together then split them 
// up into several output vectors. There are six 5-bit input vectors: 
// a, b, c, d, e, and f, for a total of 30 bits of input. 
// There are four 8-bit output vectors: w, x, y, and z, for 32 bits of output. 
// The output should be a concatenation of the input vectors followed by two 1 bits.
module top_module (
    input [4:0] a, b, c, d, e, f,
    output [7:0] w, x, y, z );//

    // Insert your code below
    // assign { ... } = { ... };

\end{lstlisting}
\vspace{-3mm}
\caption{Iter 0, Prompt with problem description, vector concatenate. }
\label{fig:vec-prompt}
\end{subfigure}

\begin{subfigure}[b]{0.98\columnwidth}
\begin{lstlisting}[language=verilog, basicstyle=\tiny\ttfamily,
linebackgroundcolor={\ifnum\value{lstnumber}>5
                \ifnum\value{lstnumber}<9
                    \color{pink}
                \fi
            \fi}
            ]
module top_module (
    input [4:0] a, b, c, d, e, f,
    output [7:0] w, x, y, z
);

wire [1:0] tmp;
assign tmp = 2'b11; 
assign {w, x, y, z, tmp} = {a, b, c, d, e, f};
endmodule
\end{lstlisting}
\vspace{-2mm}
{\footnotesize \begin{center} Iter. 0: LLM Response, Simulation Error: 25-of-26 tests fail. \end{center}}

\end{subfigure}

\begin{subfigure}[b]{0.98\columnwidth}
\begin{lstlisting}[language=verilog, basicstyle=\tiny\ttfamily,
linebackgroundcolor={\ifnum\value{lstnumber}>3
                \ifnum\value{lstnumber}<8
                    \color{pink}
                \fi
            \fi
            \ifnum\value{lstnumber}=2
                    \color{pink}
            \fi}
            ]
... SAME AS (a) TILL LINE 4
wire [29:0] concat;
assign concat = {a, b, c, d, e, f};
assign w = concat[23:16];  
assign x = concat[15:8];
assign y = concat[7:0];
assign z = {concat[5:0], 2'b11};
endmodule
\end{lstlisting}
\vspace{-2mm}
{\footnotesize \begin{center} Iter. 1: LLM Response, Simulation Error: Tests 1-2 pass. 23 tests fail.\end{center}}

\end{subfigure}

\begin{subfigure}[b]{0.98\columnwidth}
\begin{lstlisting}[language=verilog, basicstyle=\tiny\ttfamily,
linebackgroundcolor={\ifnum\value{lstnumber}>3
                \ifnum\value{lstnumber}<8
                    \color{pink}
                \fi
            \fi
            \ifnum\value{lstnumber}=2
                    \color{pink}
            \fi
            }
            ]
... SAME AS (b) TILL LINE 4
wire [29:0] concat;
assign concat = {a, b, c, d, e, f};
assign w = concat[27:20];
assign x = concat[19:12];  
assign y = concat[11:4];
assign z = {concat[3:0], 2'b11};
endmodule
\end{lstlisting}
\vspace{-2mm}
{\footnotesize \begin{center} Iter. 2: LLM Response, Simulation Error: Tests 1-10 pass and rest  fail.\end{center}}
\end{subfigure}

\begin{subfigure}[b]{0.98\columnwidth}
\begin{lstlisting}[language=verilog, basicstyle=\tiny\ttfamily,
linebackgroundcolor={\ifnum\value{lstnumber}>3
                \ifnum\value{lstnumber}<8
                    \color{pink}
                \fi
            \fi
            \ifnum\value{lstnumber}=2
                    \color{pink}
            \fi}
            ]
... SAME AS (b) TILL LINE 4
wire [29:0] concat;
assign concat = {a, b, c, d, e, f};
assign w = concat[28:21]; 
assign x = concat[20:13];
assign y = concat[12:5];
assign z = {concat[4:0], 2'b11};
endmodule
\end{lstlisting}
\vspace{-2mm}
{\footnotesize \begin{center} Iter. 3: LLM Response, Simulation Error: Tests 1-12 pass and rest fail.\end{center}}
\end{subfigure}

\begin{subfigure}[b]{0.98\columnwidth}
\begin{lstlisting}[language=verilog, basicstyle=\tiny\ttfamily,
linebackgroundcolor={\ifnum\value{lstnumber}>3
                \ifnum\value{lstnumber}<8
                    \color{green}
                \fi
            \fi
            \ifnum\value{lstnumber}=2
                    \color{green}
            \fi}
            ]
... SAME AS (b) TILL LINE 4
wire [31:0] concat;  
assign concat = {a, b, c, d, e, f, 2'b11}; 
assign w = concat[31:24];
assign x = concat[23:16]; 
assign y = concat[15:8];
assign z = concat[7:0];
endmodule
\end{lstlisting}
\vspace{-2mm}
{\footnotesize \begin{center} Iter. 4: All Tests Pass. Testbench ran successfully.\end{center} }
\end{subfigure}
\vspace{-1.0mm}
\caption{LLM: GPT-3.5-turbo, vector concat with feedback.}
\label{fig:vector-iterations}
\vspace{-5mm}
\end{figure}

\begin{figure}[t]
\centering
\begin{subfigure}[b]{0.98\columnwidth}
\begin{lstlisting}[language=verilog,basicstyle=\tiny\ttfamily]
// Assume you have a finite state machine that will identify when bytes have been correctly received when given a stream of bits. It needs to identify the start bit, wait for all 8 data bits, then verify that the stop bit was correct. If the stop bit does not appear when expected, the FSM must wait until it finds a stop bit before attempting to receive the next byte.
// Now that you have a finite state machine that can identify when bytes are correctly received in a serial bitstream, add a datapath that will output the correctly-received data byte. out_byte needs to be valid when done is 1, and is don't-care otherwise.
// Note that the serial protocol sends the least significant bit first.
// Hint: The serial bitstream needs to be shifted in one bit at a time, then read out in parallel.
module top_module(
    input clk,
    input in,
    input reset,    // Synchronous reset
    output [7:0] out_byte,
    output done
); //
	// Insert your code below
endmodule
\end{lstlisting}
\vspace{-3mm}
\caption{Iteration 0, User Prompt, Category 3: FSM serial receiver and data path}
\label{fig:fsm-prompt}
\end{subfigure}



\begin{subfigure}[b]{0.98\columnwidth}
\begin{lstlisting}[language=Verilog,basicstyle=\tiny\ttfamily,
linebackgroundcolor={%
        \ifnum\value{lstnumber}=1
                        \color{pink}
                \fi
        \ifnum\value{lstnumber}=3
                    \color{pink}
                \fi
        \ifnum\value{lstnumber}=5
                    \color{pink}
                \fi
        \ifnum\value{lstnumber}=7
                    \color{pink}
                \fi
        \ifnum\value{lstnumber}=9
                    \color{pink}
                \fi
        \ifnum\value{lstnumber}=11
                    \color{pink}
                \fi
        \ifnum\value{lstnumber}=2
                    \color{green}
                \fi
        \ifnum\value{lstnumber}=4
                    \color{green}
                \fi
        \ifnum\value{lstnumber}=6
                    \color{green}
                \fi
        \ifnum\value{lstnumber}=8
                    \color{green}
                \fi
        \ifnum\value{lstnumber}=10
                    \color{green}
                \fi
        \ifnum\value{lstnumber}=12
                    \color{green}
                \fi
    }
    ]
-   output [7:0] out_byte,
+   output reg [7:0] out_byte,
-   output done
+   output reg done
- 	assign out_byte = temp_byte;
+ 	done <= 0; // Initialize done signal
- 	counter = 0;
+ 	counter <= 0; // Use non-blocking assignment for counter in this case
- 	done = 1;
+ 	done <= 1;
- 	done = 0;
+ 	done <= 0;
\end{lstlisting}
\vspace{-3mm}
\caption{Diff b/w LLM Response from Iter. 0 and  1. Iter. 1, Compile Error: testbench failed to compile, At line 54, `done' is used as an l-value but was declared as a wire at line 6. Wires cannot be used as l-values. At line 60, `done' is used as an l-value but was declared as a wire at line 6.}
\label{fig:fsm-iter1}
\end{subfigure}

\begin{subfigure}[b]{0.98\columnwidth}
\begin{lstlisting}[language=verilog,basicstyle=\tiny\ttfamily,
linebackgroundcolor={\ifnum\value{lstnumber}=1
                    \color{pink}
            \fi
            \ifnum\value{lstnumber}>1
                \ifnum\value{lstnumber}<4
                    \color{green}
                \fi
            \fi}
            ]
- 	temp_byte <= {shift_register[7], temp_byte[7:1]};
+ 	temp_byte <= {in, temp_byte[7:1]};
+ 		out_byte <= temp_byte;
\end{lstlisting}
\vspace{-3mm}
\caption{Diff b/w LLM response from Iter. 1 and Iter. 2. Iter. 2, Simulation Error: Test 1-7,11-6,20,23-25 passed! 11 fail.}
\label{fig:fsm-iter2}
\end{subfigure}

\begin{subfigure}[b]{0.98\columnwidth}
\begin{lstlisting}[language=verilog,basicstyle=\tiny\ttfamily,
linebackgroundcolor={\ifnum\value{lstnumber}>5
                \ifnum\value{lstnumber}<10
                    \color{green}
                \fi
            \fi
            \ifnum\value{lstnumber}=1
                    \color{green}
            \fi
            \ifnum\value{lstnumber}=3
                    \color{green}
            \fi
            \ifnum\value{lstnumber}=2
                    \color{pink}
            \fi
            \ifnum\value{lstnumber}>3
                \ifnum\value{lstnumber}<6
                    \color{pink}
                \fi
            \fi
            }
            ]
+   wire start_detected;
-   start_bit = 1;
+   start_bit = shift_register[0];
-   stop_bit = 1;
-	if (stop_bit == 1) begin
+   stop_bit = shift_register[0];
+   if (stop_bit == 1 && start_detected) begin
+   // Detect if start bit is detected
+      assign start_detected = (state == RECEIVE && start_bit == 1 && counter == 0);
\end{lstlisting}
\vspace{-3mm}
\caption{Diff b/w LLM response form Iteration 2 and Iteration 3. Iteration 3, All Tests passed! Testbench ran successfully.}
\label{fig:fsm-iter3}
\end{subfigure}


\vspace{-2.0mm}
\caption{LLM: GPT-3.5-turbo, FSM serial rx (w) feedback.}
\label{fig:fsm-iters}
\vspace{-5mm}
\end{figure}

\textbf{Simulation outputs:} Test cases that pass only report ``Test $<$ID$>$ passed!'' to reduce unnecessary input token use. Here, ``ID'' refers to the test identifier from HDLBits.
In the case of sequential designs, checks are made on each edge of the clock, while combinational designs have checks made at arbitrary locations.
Test cases that fail report the full enumeration of inputs, outputs, and expected outputs given in the same order as the module definition in the design prompt. 
The total number of failing test cases is given at the end of the simulation.

\textbf{Experimental Parameters}: LLMs are probabilistic; the same prompt does not always give the same output.
Thus, each prompt was given with the same experimental setup 5 times, and the best results were used for Pass@k analysis~\cite{chen_evaluating_2021}.
The number of feedback iterations, $n$, substantially impacts the quality of generated Verilog, so we evaluated AutoChip with varying $n$, with a default $n=10$. 
Beyond this, we evaluated LLMs with their default parameters as these values are used in the normal developer-facing web interface. 

\begin{table*}[t]
\centering
\footnotesize
\caption{\small Pass@k for k=\{1,5\} (w)ith and (w/o)ithout feedback. For (w) feedback, we use $n$ iterations with either the most recent feedback (succinct) or with context retained from all iterations (full). `+GPT-4$^*$' means an ensemble using GPT-4 as a `big' LLM for addressing errors. 
}
\label{tb:results}
\begin{tabular}{l|c|c|cccc|cccc|ccccc}
\toprule
 & & \multicolumn{4}{c}{\textbf{Success (\%)}} & \multicolumn{4}{c}{\textbf{Simulation Error (\%)}} & \multicolumn{4}{c}{\textbf{Compile Error (\%)}} \\
\cmidrule(lr){4-7} \cmidrule(lr){8-11} \cmidrule(lr){12-15}
\textbf{Feedback Type} & \textbf{Metric} & \textbf{LLM} & \textbf{(w/o)} & \multicolumn{3}{c}{\textbf{(w)}} & \textbf{(w/o)} & \multicolumn{3}{c}{\textbf{(w)}} &  \textbf{(w/o)} & \multicolumn{3}{c}{\textbf{(w)}} \\
\cmidrule(lr){5-7} \cmidrule(lr){9-11} \cmidrule(lr){13-15}
& & & \textbf{n=0} & \textbf{n=1} & \textbf{n=5} & \textbf{n=10} & \textbf{n=0} & \textbf{n=1} & \textbf{n=5} & \textbf{n=10} & \textbf{n=0}& \textbf{n=1} & \textbf{n=5} & \textbf{n=10}\\
\midrule
\toprule
\multirow{10}{*}{\textbf{Succinct}} & \multirow{4}{*}{Pass@1}
&  Claude 2& 32.50 & 37.50 & 44.17 & \textbf{47.50} & 36.67 & 46.67 & 54.17 & 50.83 & 30.83 & 15.83 & 1.67 & \textbf{1.67} \\
& & GPT-3.5 (G3) & 26.45 & 30.00 & 35.00 & \textbf{37.50} & 40.50 & 50.00 & 55.83 & 57.50 & 33.06 & 20.00 & 9.17 & \textbf{5.00} \\
 & & GPT-4 & \cellcolor{gray!30}60.83 & \cellcolor{gray!30}69.16 & \cellcolor{gray!30}\textbf{81.16} & - & 19.16 & 18.33 & 12.5 & - & 20.0 & 12.5 & 7.3 & -\\
& & \textbf{G3+GPT-4$^*$} & \cellcolor{gray!20}57.05 & \cellcolor{gray!30}85.14 & \cellcolor{gray!30}\textbf{87.15} & \cellcolor{gray!30}75.18 & 20.42 & 8.84 & 9.24 & 20.96 & 22.53 & 6.02 & 3.61 & \textbf{3.86}\\
& & CodeLlama & 35.29 & \textbf{36.21} & 36.21 & 36.21 & 20.17 & 20.69 & 20.69 & 20.69 & 44.54 & 43.10 & 43.10 & \textbf{43.10} \\
& & CodeLlama+GPT-4{$^*$} & \cellcolor{gray!15}58.25 & \cellcolor{gray!20}62.53 & \cellcolor{gray!20}62.53 & \cellcolor{gray!20}62.53 & 29.21 & 31.41 & 31.41 & 31.41 & 12.52 & 6.05 & 6.05 & 6.05 \\

& & VeriGen & 27.35 & - & - & - & 12.04 & - & - & - & 60.60 & - & - & - \\
\cmidrule(lr){2-15}
 & \multirow{4}{*}{Pass@5}
&  Claude 2& 32.83 & 38.58 & 45.35 & \textbf{47.38} & 40.83 & 48.39 & 50.42 & 50.08 & 26.33 & 13.03 & 4.23 & \textbf{2.54} \\
& & GPT-3.5 (G3) & 27.27 & 31.17 & 36.00 & \textbf{39.00} & 37.69 & 49.33 & 55.50 & 54.67 & 35.04 & 19.50 & 8.50 & \textbf{6.33} \\
& & GPT-4 & \cellcolor{gray!20}63.16 & \cellcolor{gray!30}70.40 & \cellcolor{gray!30}\textbf{84.45} & - & 19.00 & 21.9 & 11.53 & - & 17.83 & 7.68 & \textbf{4.00} & -\\
& & \textbf{G3+GPT-4$^*$} & \cellcolor{gray!30}81.06 & \cellcolor{gray!30}65.39 & \cellcolor{gray!30}72.84 & \cellcolor{gray!30}\textbf{89.19} & 7.49 & 24.14 & 22.94 & 7.77 & 11.45 & 10.46 & 4.23 & \textbf{3.04} \\
& & CodeLlama & 34.29 & 35.71 & \textbf{36.59} & 36.59 & 18.82 & 21.43 & 22.47 & 22.47 & 46.89 & 42.86 & \textbf{40.94} & 40.94 \\
& & CodeLlama+GPT-4{$^*$} & \cellcolor{gray!30}70.30 & \cellcolor{gray!30}74.50 & \cellcolor{gray!30}74.75 & \cellcolor{gray!30}74.75 & 20.63 & 21.37 & 21.16 & 21.16 & 9.03 & 4.11 & 4.07 & 4.07 \\
& & VeriGen & 27.82 & - & - & - & 10.02 & - & - & - & 62.16 & - & - & - \\
\toprule
\midrule
\multirow{4}{*}{\textbf{Full}} & \multirow{2}{*}{Pass@1}
&  Claude 2& 31.67 & 33.33 & 41.23 & \textbf{42.11} & 36.67 & 56.14 & 54.39 & 54.39 & 31.67 & 10.53 & 4.39 & \textbf{3.51} \\
& & GPT-3.5 & 26.67 & 30.25 & 34.45 & \textbf{36.13} & 33.33 & 43.70 & 53.78 & 52.94 & 40.00 & 26.05 & 11.76 & \textbf{10.92} \\
\cmidrule(lr){2-15}
& \multirow{2}{*}{Pass@5}
&  Claude 2& 32.50 & 36.71 & 42.48 & \textbf{44.23} & 38.67 & 48.95 & 51.57 & 50.70 & 28.83 & 14.34 & 5.94 & \textbf{5.07} \\
& & GPT-3.5 & 28.00 & 30.47 & 34.51 & \textbf{36.36} & 35.67 & 48.82 & 56.06 & 55.39 & 38.33 & 20.71 & 9.43 & \textbf{8.25} \\
\bottomrule
\end{tabular}
\vspace{-2mm}
\end{table*}

\section{Experimental Results}
\subsection{Research Questions}
We answer the following research questions (RQs) to assess the quality of Verilog generated given the problems in~\Cref{tbl:problems-set-2}. 

\textbf{RQ1}: How do LLMs without feedback perform (i.e., zero-shot)?
\textbf{RQ2}: Does feedback improve results?
\textbf{RQ3}: Does the number of iterations of feedback impact the quality and number of correct implementations?
\textbf{RQ4}: Does retaining full context from previous iterations impact the quality?
\textbf{RQ5}: Can an ensemble of LLMs improve generation quality at reduced cost?
\textbf{RQ6}: What is the impact of iterative code generation on the cost of use and latency?

\subsection{Results}

To perform our analysis we query the chosen LLMs both without and with feedback (using AutoChip) for $n$ iterations.
Feedback is evaluated both with `succinct' feedback and `full-context' feedback.
We present results based on testbench success, simulation errors, and compilation errors.
Table~\ref{tb:results} presents best outcomes for $n$ = 0, 1, 5, and 10 using a Pass@k metric, where higher Pass@k is better performance.

\textbf{Impact of feedback}:~\Cref{tb:results} shows clear improvements in the quality of the generated Verilog with feedback. Across all LLMs, the Pass@k metrics increase substantially from the no feedback baseline ($n=0$) even with one feedback iteration (\textbf{Ans RQ1, RQ2}).

\textbf{Impact of Iterations ($n$)}: More feedback iterations continue to boost Pass@k. For example, Pass@1 for Claude 2 rises from 37.50\% at $n=1$ to 47.5\% at $n=10$ in `succint' mode, indicating that additional iterations provides more opportunity for correcting mistakes (\textbf{Ans RQ3}).

\textbf{`Full' vs `succinct' feedback}:~\Cref{tb:results} presents the proportion of code generations with success, simulation error, and compilation error with both `full' and `succinct' feedback. The results show `succinct' improves successes and reduces compilation errors as the number of iterations increases. For instance, GPT-3.5-turbo at Pass@5 has successes improve from 27.27\% to 39\% while compilation errors decline from 35.04\% to 6.33\% at 10 iterations. On the other hand, feedback with `full' context does not lead to the same consistent gains. For GPT-3.5-turbo at Pass@5, successes only increase from 28\% to 36.36\% at 10 iterations. This implies feedback containing only the most relevant errors better guides improvements over iterations. In addition, this helps reduce the total context length thus reducing the model usage cost (\textbf{Ans RQ4}). 
Though \Cref{tb:results} shows an increase in the number of simulation errors from baseline with feedback results, the number of mismatches observed during simulation with feedback drops consistently across all LLMs

\begin{figure*}[h]
    \centering
    \includegraphics[width=0.8\linewidth]{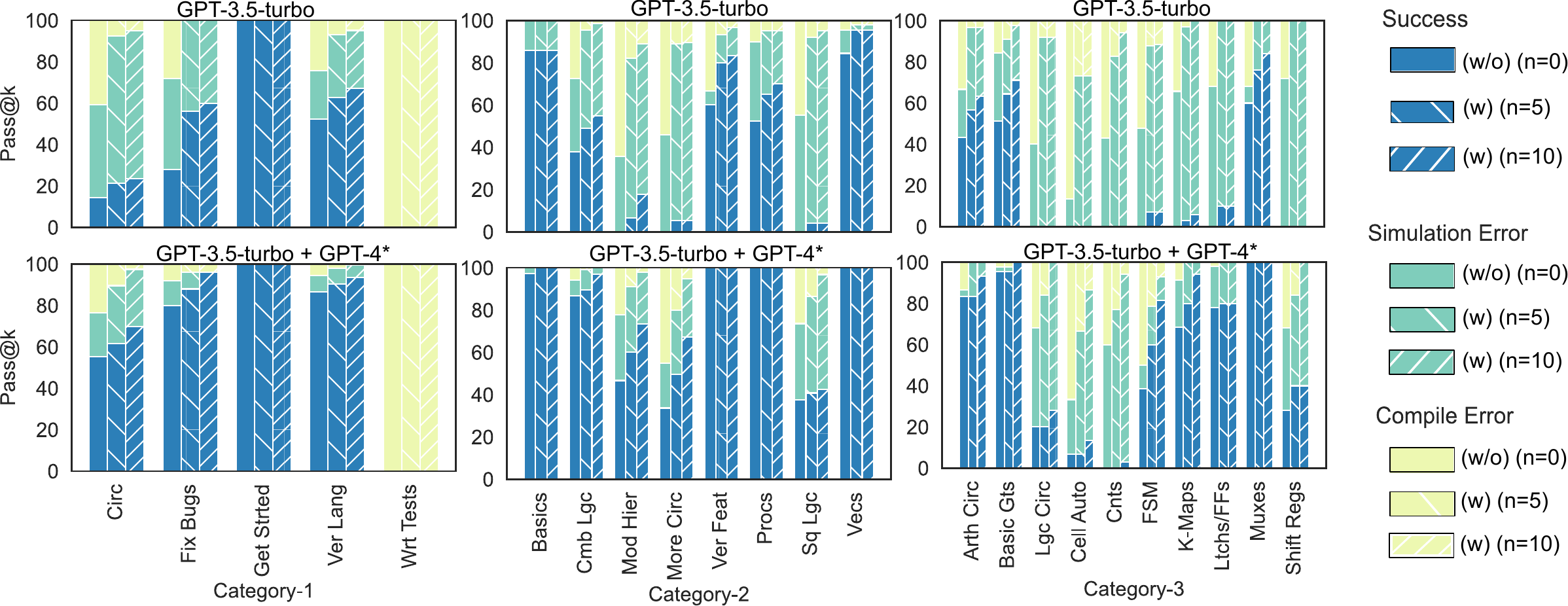}
    \vspace{-1.5mm}
    \caption{\small Top: Pass@k for best results  with GPT-3.5-turbo (w)ith and (w/o)ithout feedback, Bottom: Pass@k for best results with ensemble of GPT-3.5-turbo and GPT-4$^*$.}
    \label{fig:gpt-3.5-turbo_results_runs_5_n_15}
\end{figure*}

\textbf{Impact of LLM Ensembles}:~\Cref{fig:gpt-3.5-turbo_results_runs_5_n_15}'s top row shows the Pass@k for GPT-3.5-turbo across categories over different $n$. The bottom row shows the Pass@k for an ensemble of GPT-3.5-turbo and GPT-4. Selectively applying GPT-4 on problems where GPT-3.5 failed improved the success rate from 63\% with GPT-3.5 alone to 79\% with the ensemble. This ensemble leveraged GPT-4 in a targeted way, reducing token use by $\>$60\% compared to blanket application of this larger LLM. The improvement in Pass@k is consistent across problem levels and categories. Further, this hybrid system could solve problems that GPT-3.5 failed even after 10 iterations. 
By invoking GPT-4 upon the error, AutoChip achieved success between 20-80\% on 14 of 18 failing problems.

Leveraging GPT-4 as part of an ensemble also helps optimize cost: each input token to GPT-4 is 20$\times$ more expensive than GPT-3.5.
By only calling GPT-4 once, the total cost of the problem is greatly reduced while still giving a high percent success (\textbf{Ans RQ5, RQ6}).



\section{Discussion}

We found that prompting LLMs with iterative feedback definitively improves performance.
To understand why, we qualitatively analyze the impact of iterative feedback on improving LLM-generated Verilog code quality through two case studies: vector concatenation (\Cref{fig:vec-prompt}) and finite state machines for serial bit streams (\Cref{fig:fsm-prompt}).
In~\Cref{fig:vector-iterations}, the LLM initially struggled with vector concatenation, failing most tests; iterative feedback helped enable it to generate valid Verilog that passed tests within four rounds.
In~\Cref{fig:fsm-iters}, the LLM initially misdeclared variables and mishandled start and stop bits, causing simulation failures. Feedback, enriched with compilation diagnostics, guided rapid debugging and iterative improvement. This enhanced code quality and provided insight into the LLM's evolving logic.
Intermediate simulations enhance code quality but add to the cost.

We further found that the highest success rate was when leveraging both GPT-3.5 and GPT-4 as an ensemble; however, we see that not all types of task are doable even with this combination.
Certain classes of problem proved to consistently thwart the AutoChip framework, such as cellular automata and counters, with the testbench generation problems faring the worst, with all code unable to compile. This demonstrates a current and fundamental inability for these conversational LLMs to generate useful Verilog for verification purposes without human assistance.

Somewhat counter-intuitively, using most-recent-context feedback yields better results than full-context feedback. Likely, the LLMs are getting `confused' with the additional context that the full conversation provides. As an additional benefit, when providing the most recent context one also saves on execution cost, as the number of input tokens is consistently smaller when not sending the complete conversation for every iteration of the tool.




\section{Conclusion}

In this work we comprehensively evaluated conversational LLMs for iterative hardware development with a workflow similar to that may be undertaken by human engineers. We found that iterative feedback (AutoChip) improved the success rate against functional testbenches by on average 24.2\,\% w.r.t. baseline generation. AutoChip showed up to 89.19\% success rates (Pass@10) when using GPT-3.5 and GPT-4, suggesting that this framework provides a pathway towards the automatic design of hardware circuits.


\bibliographystyle{ACM-Reference-Format-num}
\bibliography{benhamram}

\end{document}